\documentclass[10pt,letterpaper]{article}
\usepackage[top=0.85in,left=0.75in,footskip=0.75in,marginparwidth=2in]{geometry}
\usepackage{physics}
\usepackage{nameref,hyperref}
\usepackage{graphicx}
\usepackage{color}
\usepackage{abstract,lipsum}

\usepackage[aboveskip=1pt,labelfont=bf,labelsep=period,singlelinecheck=off]{caption}

\makeatletter
\renewcommand{\@biblabel}[1]{\quad#1.}
\makeatother

\usepackage{lastpage,fancyhdr,graphicx}
\usepackage{epstopdf}
\fancyheadoffset[L]{2.25in}
\fancyfootoffset[L]{2.25in}

\usepackage{wrapfig}
\usepackage[pscoord]{eso-pic}
\usepackage[fulladjust]{marginnote}
\reversemarginpar

\begin{document}
\vspace*{0.35in}

\begin{flushleft}
{\LARGE
\textbf\newline{ Localizing an atom using Laguerre-Gaussian beams}
}
\newline
\\
Seyedeh Hamideh Kazemi\textsuperscript{1},
Mohadeseh Veisi\textsuperscript{1},
Mohammad Mahmoudi\textsuperscript{1,*}
\\
\bigskip
\textsf{1} Department of Physics, University of Zanjan, University Blvd., 45371-38791, Zanjan, Iran
\\
* mahmoudi@znu.ac.ir

\end{flushleft}
\begin{abstract}
Use of the Laguerre-Gaussian fields in an atom-light interaction makes the linewidth of the optical spectrum narrow. We exploit this fact for providing the ability to accomplish simultaneous ultra-high precision and spatial resolution atom localization in a double-$\Lambda$ atomic system; Under multi-photon resonance condition, the resolution of the localization is remarkably improved so that the atom can be localized in a region smaller than $\lambda/20 \times \lambda/20$. Most prominently, the probability of finding the atom at a particular position is always 100$\%$, when a photon with certain frequencies is absorbed or amplified. Such features are mainly dependent on radial dependence associated with the Laguerre-Gaussian fields in a spatially dependent atom-light interaction.
\end{abstract}
\vspace*{0.2in}



In recent years, many efforts have been devoted to optical techniques for precision position measurement of atoms, mainly due to the fact that these methods provide better spatial resolution. An illustrative example is the Heisenberg microscope \cite{stokes} of which the resolving power is limited by optical wavelength and allows position measurement of moving atoms as they pass through the optical fields. Further, optical methods for atom localization have been developed via an internal structure of the atomic levels. Several schemes have been proposed to achieve sub-wavelength localization of atoms \cite{5a,1a,3a,2a,4a} which has attracted wide scientific attention owing to its importance in the area of nanolithography at the Heisenberg limit \cite{11} along with its fundamental importance in laser cooling \cite{14} and trapping of atoms \cite{15,yavuz}, atom optics \cite{13}, Bose-Einstein condensation \cite{4}, measurements of the center-of-mass wave function of the moving atom \cite{6}, and fluorescence microscopy \cite{yavuz2}. Some earlier proposals of the localization include the measurement of either phase shift \cite{1a,3a,2a} or entanglement between an atom's position and its internal state \cite{4a} after passage of the atom through standing-wave field. Besides, sub-wavelength atom localization has been shown to be possible through atomic coherence, for example, detection of spontaneously emitted photons \cite{8a}, population \cite{15c}, absorption \cite{14c,mahmoudi7} and gain \cite{16c}. The main advantage of these schemes over the earlier ones is that the atom can be immediately localized during its motion, in the domain of the standing-wave field. The chief drawback of majority of the methods, however, is the fact that they yield different positions of the atom in a unit wavelength region ($\lambda$) and for a single measurement and, in favorable circumstances, the probability of finding the atom at certain positions may be $1/4$.

Two-dimensional (2D) atom localization, having broader and better prospect of application as compared with one-dimensional (1D) one, is recently studied in several papers \cite{19d,21d,20d,ding1,ding2,22d,ding,wan,jiang}. For instance, population, absorption and spontaneous emission from a driven atomic system are used to achieve a 2D localization by interacting with two orthogonal standing-wave fields. Generally, it is possible to take advantage of closed-loop atomic system for enhancing the precision of atom localization. In 2012, Ding \textit{et al} proposed a scheme for high-precision atom localization with a cyclic configuration via controlled spontaneous emission \cite{ding}. In another related study, the use of phase-dependent absorption for approaching 2D atom localization in a four-level double-$\Lambda$ system was suggested \cite{wan}. In spite of the pronounced success, there still exists a continuing need for increased precision along with higher spatial resolution of atom localization than conventional techniques can provide.

On the other hand, Laguerre-Gaussian (LG) light beams \cite{allen} having a doughnut-shaped intensity distribution and zero intensity at beam centre, have given birth to various excellent applications such as creation of a waveguide in an atomic vapour \cite{truscott}, rotating trapped microscopic particles \cite{Paterson}, and readily formation of trap split \cite{kazemi}. As mentioned before, there has been numerous number of studies related to the atom localization which adopted standing wave fields in order to have a position-dependent atom-field interaction, however, based on our knowledge none of them has been considered the LG ones. In this Letter, we present a novel scheme of atom localization using the LG fields rather than the standing-wave ones, which has allowed us to achieve 100$\%$ probability of the finding the atom at a particular position. Another prominent parameter, as far as atom localization is concerned, is spatial resolution in the probability distribution. It is already found that profile of driving fields plays a vital role in the narrowing of the linewidth of the optical spectrum \cite{hanle,ddr}. To proceed, we will show how using of the LG fields can reduce the width of the atom localization peak (dip) so that the probability distribution of the atom can be confined in $\lambda/20 \times \lambda/20$ region. It is pointed out that the effects arise mainly due to nonzero azimuthal mode index and radial dependence associated with the LG fields. The LG-induced ultra-high precision and spatial resolution localization peak (dip) may simplify a possible implementation of sub-wavelength atom localization. 

We note that our approach for atom localization has the following key advantages: 1) From experimental point of view, it is more convenient to deal with LG fields than with two-orthogonal standing wave fields. 2) Probability of finding the atom at a particular position is always 100\%, independent of the system parameters. We must reiterate the importance of the fact that maximal probability of finding an atom at certain position can be always obtained, while in the above studies involving atom localization with standing wave fields, the probability of finding the atom at the certain position can be 100\%, under favorable circumstances, only for limited parameters and limited applicability. For instance, Kou's group \cite{wan} found that the probability of detecting the atom can be 100\% merely for specific probe detunings. 3) The main feature of our suggested scheme, besides appearing just an ultra-narrow localization peak (dip) in the absorption spectra, is that spatial resolution is remarkably improved compared with the similar works \cite{20d,ding1,ding2,22d,ding,wan,jiang}. Here, it is worth comparing our scheme with a previous one about 2D atom localization in a five-level M-type atomic system, which leads to a 2D sub-half-wavelength atom localization \cite{jiang}; In addition to the obvious difference of the coupling the atom to the laser fields, there are also differences in the condition of the localization. They shown that the precision and spatial resolution of atom localization can be improved by adjusting the Rabi frequency of microwave field and the detuning of spontaneously emitted photon, which is hard to control, and in the presence of the quantum interference induced by spontaneous emission, i.e., spontaneously generated coherence (SGC). Noting that rigorous condition of the non-orthogonal dipole moments and also near degenerated states are barely met in real atomic systems. Another merit of our approach is that it is based on the probe-absorption measurement which is much easier to carry out in a laboratory compared to the measurement of spontaneous emission. These advantages may provide a possibility to observe a 2D sub-wavelength atom localization in the experiment.

\begin{figure}[t]
\centering
\includegraphics[width=7cm]{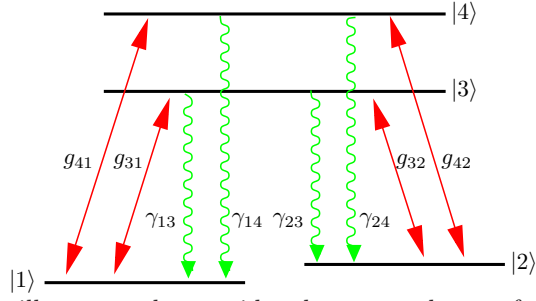}
\caption{ The figure illustrates the considered energy scheme of the four-level double-$\Lambda$ atomic system in which wavy lines show the spontaneous decays from the excited states. }
\label{fig1}
\end{figure}

As shown in Fig.~\ref{fig1}, we consider a closed-loop four-level double-$\Lambda$ atomic system interacting with four coherent fields, whose level structure consists of two metastable lower states $\vert 1\rangle$ and $\vert 2\rangle$ plus two excited states $\vert 3\rangle$ and $\vert 4\rangle$. All of the allowed electric dipole transitions, i.e., $ \vert 1\rangle - \vert 3\rangle$, $\vert 2\rangle - \vert 3\rangle$, $\vert 1\rangle - \vert 4\rangle$ and $\vert 2\rangle - \vert 4\rangle$ are driven by the laser fields; The ground level $\vert 1\rangle$ is coupled to the excited levels $\vert 3\rangle$ and $\vert 4\rangle$ by control and probe fields $E_{31}$ and $E_{41}$ with carrier frequencies $\omega_{31}$ and $\omega_{41}$, respectively. Two coupling fields $E_{32}(x,y)$ and $E_{42}(x,y)$ with carrier frequencies $\omega_{32}$ and $\omega_{42}$ couple the excited states to another lower state $\vert 2\rangle$. The spontaneous decay rates on the dipole-allowed transitions are denoted by $\gamma_{13}$, $ \gamma_{23}$, $ \gamma_{14}$, and $ \gamma_{24}$. 

In our suggested scheme for 2D atom localization, two fields $E_{31}$ and $E_{41}$ propagate in the $y$ direction and the coupling fields $E_{32}$ and $E_{42}$ have LG profile. As an atom passes through the LG fields with high enough velocity and the interaction with the fields does not effect its motion along the $z$ direction, atomic motion treated as a classical case. However, the velocity of the atom along the $x(y)$ direction is very small and center-of-mass position of the atom is nearly constant during the interaction time. Thus, we can neglect the kinetic energy in the Hamiltonian by applying the Raman-Nath approximation. Under the electric-dipole and rotating-wave approximations, the Hamiltonian for the system can be written as 
\begin{eqnarray}\label{hamil}
H_{int} &=&-\hbar [ \, g_{31} e^{i(\Delta_{31} t +\phi_{31})} \vert 1\rangle  \langle 3 \vert + g_{41} e^{i(\Delta_{41} t +\phi_{41})} \vert 1\rangle  \langle 4 \vert \\ \nonumber 
 &+& g_{42}(x,y) e^{i(\Delta_{42} t +\phi_{42})} \vert 2\rangle  \langle 4 \vert + g_{32}(x,y) e^{i(\Delta_{32} t +\phi_{32})} \vert 2\rangle  \langle 3 \vert + \mathrm{H.c.}].
\end{eqnarray}
Here, H.c. corresponds to the Hermitian conjugate of the terms explicitly written in the Hamiltonian. The parameter $\Delta_{ij}= \omega_{ij}-\bar{\omega}_{ij}$ denotes the detuning of the laser field from the corresponding transition, with $\bar{\omega}_{ij}$ being as the frequency of the transition $ \vert i\rangle \leftrightarrow \vert j\rangle$ ($i \in \lbrace 3,4 \rbrace$ and $j \in \lbrace 1,2 \rbrace$). Moreover, $\phi_{ij}$ is the phase of the laser field interacting with the corresponding transition. The general expression for a Rabi frequency can be written as $g_{ij}=(\vec{\mu}_{ij}. \vec{E}_{ij})/{\hbar}$, where $\vec{\mu}_{ij}$ and $\vec{E}_{ij}$ are the atomic dipole moment of the corresponding transition and peak amplitude of the field, respectively.
\begin{figure*}[!ht]
\centerline{\includegraphics[width=13cm]{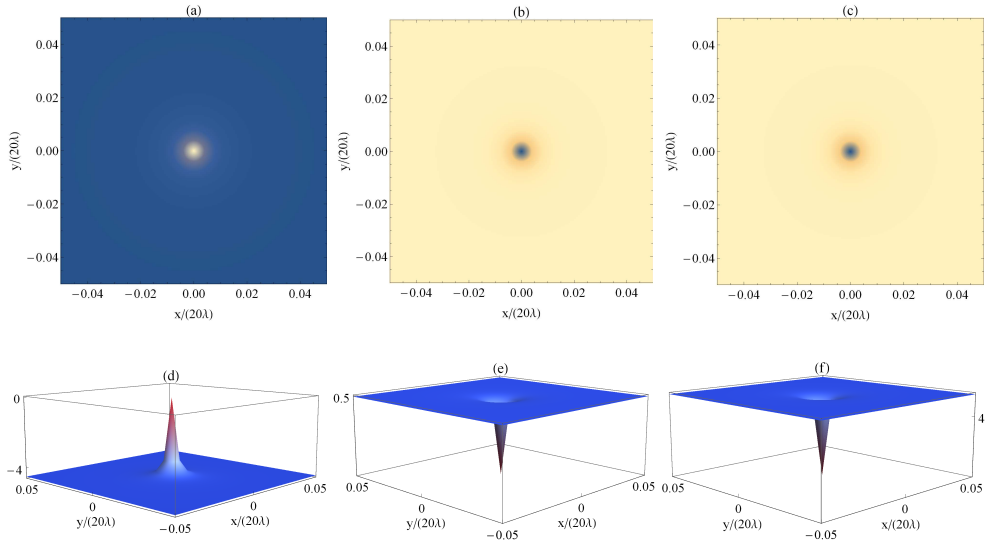}}
\caption{The filter function $\mathcal{F}(x,y)$, directly describing the conditional position probability distribution, versus the normalized position ($x/20\lambda$,$\,y/20\lambda$) for $\Delta_{31} = \Delta_{32} = \Delta_{42} = 0$, $g_{31}=0.01 \gamma$, $g_{41}= 10^{-4}\gamma$, $g^{'}_{32}=g^{'}_{42}= \gamma$ and $w= 1\,\mu$m. (a) $\phi=0$, (b) $\phi=\pi/2$ and (c) $\phi=\pi$. (d), (e) and (f) depict three-dimension plot associated with (a), (b) and (c), respectively. }
\label{fig2}
\end{figure*}

LG beam (LG$_{p}^{l}$) defines a solution of the paraxial wave function in a cylindrical coordinate which its indices $l$ and $p$ are the number of times the phase completes 2$\pi$ on a closed loop around the axis of propagation and the number of radial node for radius $r > 0$, respectively \cite{hanle,allen2}. In this Letter, the azimuthal mode index ($l$) associated with the LG beam is taken as $l=1$ and we also assume that the radial mode index is $p=0$. So, the Rabi frequancy for the LG beams (LG$_{0}^{1}$) is written as  
\begin{equation}\label{lg}
g_{i2} (x,y)=  g^{'}_{i2} \,e^{-i \varphi} \, \dfrac{ \sqrt{x^2+y^2}}{w} \exp \left(  -\dfrac{x^2+y^2}{w^2} \right),
\end{equation}
where $g^{'}_{i2}$ is the Rabi frequency constant and $\varphi=\arctan(y/x)$. Note that we ignore the $z$ dependence in the region $z\ll \,$Rayleigh length $(z_{R})$ and $w$ represents the beam waist for a beam width of $w_0$ ($w \approx w_{0} $ for $z\ll z_{R}$) \cite{allen2}.


Thy dynamic of the system can be describes by the density-matrix equations of motion which have been worked out in Ref. \cite{mahmoudi}. Note that, by choosing multi-photon resonance condition, $ \Delta_{32}+\Delta_{41}-\Delta_{31}-\Delta_{42}=0 $, the case in which the coefficients of the density-matrix equations do not have an explicit time dependence exponential factor and thus a stationary steady-state in the long-time limit can be found.

Because of position-dependent Rabi frequencies $g_{32}$ and $g_{42}$, the Hamiltonain and the dynamics of the system are position dependent and in principle, the 2D position information of the atom can be extracted via the susceptibility at the probe field frequency. The linear susceptibility of the probe field can be written as $\chi = N \eta_{p} \rho_{41}/(\epsilon_0 E_{31}) $ with $N$, $\eta_{p}$ and $\rho_{41}$ being as, respectively, atom number density in the medium, the probe transition dipole moment and the atomic coherence \cite{scully}. Real and imaginary parts of the susceptibility correspond to the dispersion and absorption, respectively. Notice that, throughout the discussion, for simplicity we have written the susceptibility as $\Im[\chi]= (N \eta_{p} / \epsilon_{0})  \mathcal{F}(x,y) $ via definition a filter function $ \mathcal{F}(x,y)$ which determines the position probability distribution of the atom \cite{22d,wan}. 

Using the phase-dependent absorption \cite{wan} or the SGC \cite{jiang}, as mentioned above, the number of the localization peaks in a unit wavelength domain of the classical standing-wave fields can be reduced to $1$ and the detection probability at a certain position can be improved to 100\%. Keeping in mind their innovative ideas, the main question here is: Is there any way to readily high precision position measurement of the moving atom? In the following, we show how we can have just an ultra-narrow localization peak (dip) as well as the improved spatial resolution without encountering difficulty in realizing the SGC or controlling detunings. As the probe absorption spectrum depends on $x,y$ from the coupling fields, it is possible to extract the 2D position information of the atom when it is passing through the fields via measuring the probe absorption or gain. Our results are represented in scaled quantities to obtain the best possible comparison with other localization schemes; The positions are divided by $20\lambda$ with a typical value of the wavelength $\lambda=253 \,nm$.

In Fig.~\ref{fig2}, we analyze the filter function $\mathcal{F}(x,y)$ which shows the atomic position probability distribution and is obtained by numerically solving the density-matrix equations in the long-time limit. The common parameters are chosen as $ \gamma_{13}=  \gamma_{14}=  \gamma_{23}= \gamma_{24}=\gamma$, $g_{31}=0.01 \gamma$, $g_{41}= 10^{-4}\gamma$, and $g^{'}_{32}=g^{'}_{42}= \gamma$. We also assume that the carrier frequencies of the fields satisfy the multi-photon
resonance condition, i.e., $\Delta_{31} = \Delta_{32} = \Delta_{42} = \Delta= 0$. Further, all calculations are performed by assuming that the beam width of the each LG field is given by $w= 1\, \mu$m. Due to the closed-loop configuration, the absorption and gain properties of the system depend on the relative phase of the fields: $\phi =\phi_{32}+\phi_{41}-\phi_{31}-\phi_{42}$. According to the definition of the filter function, $\mathcal{F}(x,y)>0$  means that the probe field is absorbed by the atom, while on the condition $\mathcal{F}(x,y)<0$ corresponds to the amplification. In order to describe clearly the localization properties, three-dimensional plots of the filter function are presented in Figs.~2(d), 2(e) and 2(f) corresponding to the Figs.~2(a), 2(b) and 2(c), respectively. 

First of all, we consider the case of $\phi=0$ and plot the filter function versus the normalized position ($x/20\lambda$,$\,y/20\lambda$). As can be seen in Fig.~\ref{fig2}(a) and \ref{fig2}(d), an ultra-narrow localization peak with a full width at half-maximum (FWHM) of less than 0.004 $\lambda$ (about 1 nm) is found and the probability of detecting the atom is 100\%, determined by the gain of the probe field. On the condition of $\phi=\pi /2$, the filter function has a narrow dip (see Fig.~\ref{fig2}(b) and \ref{fig2}(e)). For the case of $\phi=\pi$, it takes the forms similar to that for the previous case: presenting narrow dip, but with a significantly larger depth (Fig.~\ref{fig2}(c) and \ref{fig2}(f)). That to say, when a probe photon is absorbed, we can sure that the atom is passing through the fields from the position determined by the circle in the figure and the probability of the finding the atom at such position also is 100\% which is greatly improved comparing with previous schemes \cite{21d,20d,ding1,ding2,22d,wan}. It is also obvious that the atom is located in a region smaller than $\lambda/20 \times \lambda/20$. Indeed, our suggested scheme improves previous results that atom is confined in $\lambda/10 \times \lambda/10$ \cite{jiang} region or wider ones \cite{20d,ding1,ding2,22d,ding,wan}. It is imperative to point out that the ultra-high precision and spatial resolution atom localization is independent of choosing the detunings in such a way that similar behavior is found for the case of nonzero detunings, but peak with smaller height (dip with smaller depth) for the case of $\Delta > \gamma$.  

By solving the equations of motion in steady-state, we can derive analytical solutions for the probe susceptibility and, consequently, the filter function. In the particular case of $\Delta= 0$ and $g_{32}(x,y)=g_{42}(x,y)=g(x,y)$, the expression for the $\rho_{41}$ is given as
\begin{eqnarray}
 \Im [\rho_{41}(x,y)]&=& \dfrac{-\gamma}{2 D^2} ( D \, g_{31}\,g^2(x,y)\, \Im[i\,  e^{i \phi}] \\ \nonumber
&-&  g_{41} \, \gamma ^2  \, g^2(x,y)\, (g_{31}^2 +2\, g^2(x,y)) 
+ g_{31}^2 \, g^{\ast}_{41} \,g^4(x,y)  \Im[i \, e^{2i \phi}]), \label{eq4.1}
\end{eqnarray}
where $D=g_{31}^2 \, g^2(x,y)+ \gamma ^2 \,(g_{31}^2 +2 g^2(x,y)) $. Theses terms have a simple interpretation in terms of the physical processes \cite{mahmoudi}. The first one, proportional to $g_{31} \, g_{32}^{\ast}\,g_{42}$, corresponds to the closed interaction loop and also a scattering of the driving field modes to the probe field mode. The second component proportional to $g_{41}$ represents a direct scattering of the probe field off of the probe transition. The last term proportional to $g_{41}^{\ast}$ can be interpreted as a counter-rotating propagation. The expression also allows us to analyze the atom localization: The height of the peak (the depth of the dip) is  remarkably affected by the relative phase of the applied fields in such a way that in the case of $\phi=\pi/2$ and our parameter ranges, we have only two small last terms in Eq.~3 and the depth is much smaller than that of $\phi=\pi$. The filter function for the case of $\phi=\pi$ here also is approximately reverse to that of the $\phi=0$, as two last terms are very small. Furthermore, the optimal results, i.e., high precision and high resolution atom localization, can be achieved for the LG fields with equal azimuthal mode index ($l$). One the other hand, the filter function for such cases becomes independent of the phase factor ($e^{-i l \varphi}$) of the coupling fields, as predicted from Eq.~3. Thus, by attending to these points, it can be claimed that the radial dependence, not the phase factor, associated with the LG fields brings about the observed ultra-high precision and spatial resolution atom localization.

From Eq.~3, we then obtain an analytical expression for the FWHM of the appeared peak (dip) along the x(y) axis. As shown above, the depth of the dip for the case $\phi=\pi/2$ is smaller than that of $\phi=\pi$, and subsequently we have a more effective atom localization for the latter case, so we will only concentrate on the case $\phi=\pi$. In the weak probe regime and for a moderate control field ($g_{41} < \gamma$), the FWHM can be derived from expression
\begin{equation}
\mathrm{FWHM}=  \sqrt{ \dfrac{ 2 \gamma ^2 \, w^2 \, g_{31}^2 }{ 2 \gamma ^2 \,g^{'2} +16 \gamma ^2 g_{31}^2 + g_{31}^2 \, g^{'2}} }\, \, ,
\label{eq5}
\end{equation}
where $g^{'}_{32}=g^{'}_{42}=g^{'}$. One can see from Eq.~\ref{eq5} that a narrower localization dip can be attained by decreasing the beam width of the LG fields ($w$) and the intensity of the control field, as well as increasing the Rabi frequency constant of the coupling fields ($g^{'}$). It turns out that an interesting parameter range for the present study is given by $ g_{31} \ll g^{'}$; in this limit, the half width of the dip along the x(y) axis is determined by the ratio $w  / g^{'} $, hence, the high-precision localization is obtained by means of decreasing this ratio.


In summary, this Letter has proposed and analyzed a novel atom localization scheme permitting simultaneous ultra-high precision and spatial resolution which takes advantage of the sharp spectra induced by the LG fields. Under multi-photon resonance condition, the atom can be localized in a region smaller than $\lambda/20 \times \lambda/20$ and probability of the finding the atom at a particular position is always 100$\%$. Taking into account that the spatial resolution as well as the precision of the atom localization are significantly improved over those provided by conventional localization techniques. Such features are mainly attributed to radial dependence associated with the LG fields in a spatially dependent atom-light interaction.

\bibliographystyle{}

\end{document}